\documentstyle[aps,twocolumn,floats,prl, epsfig]{revtex}

\newcommand {\phidot}{\dot{\phi}} 

\newcommand {\mpl}{m_{\scriptscriptstyle{PL}}}
\newcommand {\ts}{\textstyle}
\newcommand {\ds}{\displaystyle}

\def\GeV{\,{\rm GeV}}

\def\la{\mathrel{\mathpalette\fun <}}
\def\ga{\mathrel{\mathpalette\fun >}}
\def\fun#1#2{\lower3.6pt\vbox{\baselineskip0pt\lineskip.9pt
  \ialign{$\mathsurround=0pt#1\hfil##\hfil$\crcr#2\crcr\sim\crcr}}}

\begin{document}
\input epsf
\twocolumn[\hsize\textwidth\columnwidth\hsize\csname
@twocolumnfalse\endcsname

\title{On the degree of scale invariance of inflationary perturbations}
\vskip.25in
\author{Dragan Huterer$^1$ and Michael S. Turner$^{1,2,3}$}
\vskip.25in

\address{\it $^1$Department of Physics,
The University of Chicago, Chicago, IL~~60637-1433} %

\address{\it $^2$Department of Astronomy \& Astrophysics \\
Enrico Fermi Institute, The University of Chicago, Chicago, IL~~60637-1433}

\address {\it $^3$NASA/Fermilab Astrophysics Center\\
Fermi National Accelerator Laboratory, Batavia, IL~~60510-0500}

\maketitle
\begin{abstract}

Many, if not most, inflationary models predict the power-law
index of the spectrum of density perturbations is close to one, though
not precisely equal to one, $|n-1| \sim {\cal O}(0.1)$, implying that
the spectrum of density perturbations
is nearly, but not exactly, scale invariant.  Some
models allow $n$ to be significantly less than one
($n\sim 0.7$); a spectral index significantly greater than one
is more difficult to achieve.  We show that $n\approx 1$ is
a consequence of the slow-roll conditions for inflation and
``naturalness,'' and thus is a generic prediction of inflation.
We discuss what is required to deviate significantly
from scale invariance, and then show, by explicit construction,
the existence of smooth potentials that satisfy all the conditions for
successful inflation and give $n$ as large as 2.

\end{abstract}

\vskip 1.0truecm
]

\section{Introduction}\label{sec-intro}

Inflation generates adiabatic density perturbations that can seed
the formation of structure in the Universe.  They arise from
quantum fluctuations in the field that drives inflation and are
stretched to astrophysical size by the enormous growth of the
scale factor during inflation \cite{scalar}.  The magnitude
of these perturbations was recognized early on to be
important in constraining inflationary models.  The
nearly scale-invariant value for the scalar spectral index,
$n\approx 1$, is considered to be one of the three principal
predictions of inflation, and the deviation of $n$ from unity is
an important probe of the underlying dynamics of inflation
\cite{Turner_10things}.

The advantage of scale-invariant primordial density perturbations
was first spelled out nearly three decades ago \cite{Harrison,Zeldovich}:
any other spectrum, in the absence of a long-wavelength or short-wavelength
cutoff, will have excessively large perturbations on small scales or
large scales.\footnote{Inflation provides a natural cutoff on
comoving scales smaller than $\sim$1\,km, the horizon size at
the end of inflation; perturbations on scales larger than the
present horizon will not be important until long into the future.
Thus, for inflation exact scale invariance is not necessary to avoid problems
with excessively large perturbations.}
Even though inflation provided the first realization of such
a spectrum, long before inflation many cosmologists considered
the scale-invariant spectrum to be the only sensible one.  For this
reason, the inflationary prediction of a deviation from scale
invariance -- even if small -- becomes all the more important.

One of the pioneering papers on inflationary fluctuations
\cite{bst} emphasized that the fluctuations were not precisely
scale-invariant; the first quantitative discussion
followed a year later \cite{Steinh_Turner}.
The COBE DMR detection of CBR anisotropy awakened the inflationary
community to the testability of the inflationary density-perturbation
prediction.  The connection between $(n-1)$ and the underlying
inflationary potential was pointed out soon thereafter
\cite{davisetal,Liddle_Lyth}, and the possibility of reconstructing
the inflationary potential from measurements of CBR anisotropy
began being discussed \cite{recon}.  It is now quite clear that
the degree of deviation from scalar invariance is an important
test and probe of inflation.

Particular inflationary potentials and the values of $n$ they predict
have been widely discussed in literature (see e.g.,
Refs.~\cite{Turner93,Lyth96}).  Lyth and Riotto \cite{Lyth96}, for example,
remark that many inflationary potentials can be written
in the form $V(\phi )=V_0(1\pm \mu \phi^p)$ (in the interval
relevant for inflation), and conclude that virtually all
potentials of this form give $0.84<n<0.98$ or $1.04<n<1.16$
(also see Ref.~\cite{Steinh_Turner}).  Experimental limits on $n$,
derived from CBR anisotropy
measurements, are not yet very stringent, $0.7<n<1.2$
\cite{White_etal,Tegmark}.  Even the stronger bound claimed by
Bond and Jaffe \cite{bond_etal}, $n=0.95\pm 0.06$, falls far short
of the potential of future CBR experiments
(e.g., the MAP and Planck satellites),
$\sigma_n \sim 0.01$ \cite{Eisenstein_etal}.

The purpose of our paper is to discuss the general issue of
the deviation from scale invariance, and to explain why scale invariance is
a generic feature of inflation.  In so doing, we will take
a very agnostic approach to models.  In view of our
lack of knowledge about physics of the scalar sector and
of the inflationary-energy scale, this seems justified.
As we show, the slow-roll conditions necessary for
inflation are closely related to the possible deviation from
scale invariance.  To illustrate what must be done to
achieve significant deviation from scale invariance,
we discuss models based upon smooth potentials where $n$ is
much smaller than and much larger than unity.

\section{Why inflationary perturbations are nearly
scale invariant}\label{sec-ngg1}

The equations governing inflation are well known \cite{kt}
\begin{eqnarray}
\ddot\phi + 3H\dot\phi + V^\prime (\phi ) & = & 0 \\
H^2 \equiv \left({\dot a\over a}\right)^2
        & = & {8\pi \over 3\mpl^2}\,\left[V(\phi )
        + {1\over 2}\dot\phi^2\right] \\
        N \equiv \ln (a_f/a_i)
        & = & \int_{\phi_i}^{\phi_f}Hdt \\
\delta_H^2 (k) & \simeq & V^3/V^{\prime 2} \propto k^{n-1},
\end{eqnarray}
where $a(t)$ is the cosmic scale factor,
derivatives with respect to the field $\phi$ are denoted
by prime, and derivatives with respect to time by overdot.
The quantity $\delta_H$ is the post-inflation horizon-crossing
amplitude of the density perturbation, which, if the perturbations
are not precisely scale invariant is a function of comoving wavenumber $k$.
(The dimensionless amplitude $\delta_H$ also corresponds to
the dimensionless amplitude of the fluctuations in the gravitational
potential.)

In computing the density perturbations,
the value of the potential and its first derivative are
evaluated when the scale $k$ crossed outside the horizon
during inflation.   Because both $V$ and $V^\prime$
can vary, $\delta_H^2 \propto k^{n-1}$ in general depends
upon scale; exact scale-invariance corresponds to
$n=1$.  For most models, $\delta_H^2$ is not a true power
law, but rather $n$ varies slowly with scale, typically
$|dn/d\ln k|\leq 10^{-3}$ \cite{Kosowsky_Turner};
in fact, both $n$ and $dn/d\ln k$
are measurable cosmological parameters and can
provide important information about the potential.

In the slow-roll approximation the $\ddot\phi$ term
is neglected in the equation of motion for $\phi$ and the
kinetic term is neglected in the Friedmann equation \cite{Steinh_Turner,kt}:
\begin{eqnarray}
\dot\phi & \simeq & {V^\prime \over 3H} \\
N & \simeq & -{8\pi\over \mpl} \int_{\phi_i}^{\phi_f}
         {d\phi \over x(\phi )}. \\
\end{eqnarray}
The power-law index $n$ is given by \cite{Liddle_Lyth,Turner93}
\begin{equation}
(n-1) =  -\frac{x_{60}^2}{8 \pi} +
\frac{\mpl x_{60}'}{4 \pi}, \label{n-1}\\
\end{equation}
where $x(\phi ) \equiv \mpl V^\prime (\phi ) /V(\phi )$ measures
the steepness of the potential and $x^\prime =dx/d\phi$ measures the
change in steepness.  (Higher-order corrections are discussed
and the next correction is given in Ref.~\cite{LT}.)
The subscript ``60'' indicates that these parameters are evaluated
roughly 60 e-folds before the end of inflation, when the scales
relevant for structure formation crossed outside the horizon.

Deviation from scale invariance is a generic prediction
since the inflationary potential cannot be absolutely flat,
and it is controlled by the steepness and the change in steepness
of the potential.  Significant
deviation from scale invariance requires a steep potential or
one whose steepness changes rapidly.  Further,
Eq.~(\ref{n-1}) immediately hints that it is easier to make models
with a ``red spectrum'' ($n < 1$), than with a ``blue spectrum''
($n > 1$), because the first term in Eq.~(\ref{n-1})
is manifestly negative, while the second term
can be of either sign. In addition, $x_{60}^2/8 \pi$ is usually larger
in absolute value than $\mpl x_{60}'/4\pi$.

The two conditions on the potential needed to ensure the validity
of the slow-roll approximation are (see e.g., Refs.~\cite{Steinh_Turner,kt}):
\begin{eqnarray}
\mpl V^{\prime}/V = x & \la & \sqrt{48\pi} \\
\mpl^2 V^{\prime\prime}/V &\la & 24\pi.
\end{eqnarray}
Note that the first slow-roll condition constrains the first
term in the expression for $(n-1)$, and
the second slow-roll condition constrains
the second term since, $\mpl x^\prime = \mpl^2  V^{\prime\prime}/V 
 -x^2$.  

A model that can give $n$ significantly less than 1 is power-law inflation
\cite{Abbott_Wise,Lucchin_Mattarese} (there are other models too
\cite{Steinh_Turner,Freese_etal}).
It also illustrates the tension
between sufficient inflation and large deviation from scale invariance.
The potential for power-law inflation is exponential,
\begin{equation} 
V=V_0 \exp(-\beta \phi/\mpl), 
\end{equation} 
the scale factor of the Universe evolves according to a power law
\begin{equation} 
a(t) \propto   t^{16 \pi/\beta^2} \equiv  t^p
\qquad \mbox{with} \qquad p\equiv  16 \pi/\beta^2,
\end{equation}
and
\begin{equation}
\dot\phi = \sqrt{p\over 4\pi}{\mpl \over t}.
\end{equation}
Further, $n$ can be calculated exactly in the
case of power-law inflation \cite{Lyth_Stewart_92}
\begin{equation}
(n-1) = {2\over 1-p}\  \rightarrow\  -{2\over p}\ \  {\rm (slow-roll\ limit)}.
\end{equation}

For this potential $x = -\beta$, $x'=0$ (constant steepness), and the slow-roll
constraint implies $|\beta| \la 7$, or $p \ga 1$.  This is
not very constraining as $p>1$ is required for the superluminal
expansion necessary for inflation \cite{htw}.  The
quantitative requirement of sufficient inflation to solve the
horizon problem
and a safe return to a radiation-dominated Universe before big-bang
nucleosynthesis (reheat temperature $T_{\rm RH} \gg 1\,$MeV and
reheat age $t_{\rm RH} \ll 1\,$sec) and baryogenesis ($T_{\rm RH}
> 1\,$TeV and $t_{\rm RH} < 10^{-12}\,$sec) restricts $p$
more seriously.

In particular, the amount of inflation is depends upon when inflation ends:
\begin{equation}
N = -{8\pi \over \mpl}\int^{\phi_f}_{\phi_i}
        \,{d\phi \over x(\phi )} = p\ln (H_i/H_f),
\end{equation}
where $H_i = p/t_i$ and $H_f=p/t_f$.  The number of e-folds $N$ required
to solve the horizon problem (i.e., expand a Hubble-sized patch
at the beginning of inflation to comoving size larger than the
present Hubble volume) is approximately 60, but
depends upon $H_i$ and $H_f$ if $p$ is not $\gg 1$ (see e.g.,
Ref.~\cite{kt}):
\begin{equation}
N > 74 + \ln (H_i/H_f) +{1\over 2}\ln (H_f/\mpl ).
\end{equation}

Bringing everything together, the constraint to $p$ is
\begin{equation}
p > 1 + {74\over \ln (H_i /H_f)} + {1\over 2} {\ln (H_f/\mpl )
        \over \ln (H_i/H_f)}.
\end{equation}
Based upon the gravity-wave contribution to CBR anisotropy
$H_i$ must be less than about $10^{-5}\mpl$ and the baryogenesis
constraint implies $H_f \ga (1\,{\rm TeV})^2/\mpl \sim
10^{-32}\mpl$.  Since reheating is not expected to be very
efficient and baryogenesis may require a temperature much
greater than $1\,$TeV (if it involves GUT, rather than electroweak,
physics), we can safely say that $H_f \gg 10^{-32}\mpl$.
Thus, sufficient inflation and safe return to
a radiation-dominated Universe before baryogenesis requires:
\begin{eqnarray}
p & \gg & 2 \\
(1-n) & \ll & 2.
\end{eqnarray}
Even insisting that $H_f \ga (10^{13}\GeV )^2/\mpl$, a typical
inflation scale, only leads to $p\ga 5$ and $n\ga 0.5$,
which is still a large deviation from scale invariance.

While the exponential potential allows a very large deviation from
$n=1$, it illustrates the tension between achieving sufficient inflation
and large deviation from scale invariance: because $(1-n) = 2/(p-1)$,
large deviation from scale invariance implies a slow, prolonged
inflation, $\ln (t_f/t_i) \simeq N(1-n)/2$, with the change in the
inflaton field being many times the Planck mass, $\Delta \phi \simeq N
\sqrt{(1-n)/(8\pi)}\, \mpl \gg \mpl$.  Other models also exhibit this
tension: For example, for the potential $V(\phi ) = V_0-m^2\phi^2/2 +
\lambda\phi^4/4$, the lower limit to $n$ is set by the condition of
sufficient inflation \cite{Steinh_Turner}.

Achieving $n$ significantly greater 1 provides a different challenge
since the first term in the equation for $(n-1)$ is negative and
the work must be done by the change-in-steepness term, $\mpl x^\prime /4\pi$.
To see the difficulty of doing so, let us assume that
we can expand the slow-roll parameter $x(\phi)$ around a point 
$\phi_*$ in the slow-roll region:
\begin{equation} 
x(\phi)\approx x_{*}+x_{*}'(\phi-\phi_{*}).  \label{Taylor}
\end{equation}  
This expression holds
for potentials whose steepness does not change much in the
slow-roll  region.  $N$ can now be evaluated explicitly:
\begin{eqnarray}
N = -\frac{8 \pi}{\mpl}\int_{\phi_i}^{\phi_f} \frac{d\phi}{x(\phi)}
=  \frac{8 \pi}{x_{60}'\mpl}\ln{\left ( \frac{x_i}{x_f}\right )}, \label{N}
\end{eqnarray}
where $x_i$ and $x_f$ are understood to have been evaluated according
to expression (\ref{Taylor}).  Combining expressions
(\ref{N}) and (\ref{n-1}), we get
\begin{equation}
n-1=\frac{2}{N}\ln{\left (\frac{x_i}{x_f}\right)} - \frac{x_{60}^2}{8 \pi},
\label{hard}
\end{equation}
and the difficulty of obtaining large $n-1$ is now more transparent. For
example, to get $n\approx 1.5$ with $N\geq 60$ we need $\ln(x_i/x_f)
> 15$ -- more, if $x_{60}^2/8 \pi$ is not negligible.  Not only does
such a large change seem unnatural, but it probably invalidates
the expansion in Eq.~(\ref{Taylor}).

Note, Eq.~(\ref{hard}) (and others below) make it appear that
$(n-1)$ depends directly upon the amount of inflation.  This is
not really the case, because $N$ is the number of e-folds that
occur during the time $x$ evolves from $x_i$ to $x_f$.  In
relating $(n-1)$ to properties of the potential it is probably
most useful to set $N=60$, and further to expand $x(\phi)$ around
$\phi_{60}$, the era relevant to creating our present Hubble
volume. Therefore, we choose $\phi_i=\phi_{*}=\phi_{60}$.

Now further specialize to the case where
$x_{60}^2/8\pi \ll |\mpl x_{60}^\prime |/4\pi$ and $|x_{60}|\gg
|x_{60}'\Delta\phi|$, where $\Delta \phi = \phi_f-\phi_i$.
Here we have explicitly assumed that the change in the steepness of
the potential is small.  It now follows that
\begin{eqnarray}
N & \simeq & {8\pi \over \mpl }{\left | \Delta \phi \over
x_{60}\right |}  \\
(n-1) & \simeq & {2\over N}
        {\left |\Delta \phi \over x_{60}\right |} \,
       x_{60}^\prime < {2\over N}
\end{eqnarray}
(note that $\Delta \phi$ and $x_{60}$ are of opposite sign). 
Thus, we get a very strong constraint on $n$ in this case,
$(n-1)< 0.04$, and learn that to achieve $n$ significantly
greater than unity, the scalar field must change by much
more than $\mpl$.

One well-known class of inflationary models that gives $n\geq 1$ is
hybrid inflation \cite{Linde}; in the slow-roll region, $ V(\phi )
\simeq V_0(1+ \mu \phi^2)$.  In these models,
\begin{eqnarray}
N & \simeq & {4\pi \over \mu \,\mpl^2} \ln (\phi_i/\phi_f) \\
(n-1) & \simeq & {\mpl x^\prime \over 4\pi} = {\mu\, \mpl^2\over 2\pi}
        \simeq {2\over N} \ln (\phi_i/\phi_f).
\end{eqnarray}
Thus, $n$ significantly larger than 1 can be achieved,
albeit at the expense of an exponentially long
roll, $\phi_i/\phi_f=\exp [N(n-1)/2]$.
However, $\phi_f$ may
not be arbitrarily small here -- in fact, the smallest value it can take
in the semi-classical approximation is equal to the magnitude of quantum
fluctuations of the field, $H/2\pi$ (this is further discussed in the
next section).  This constraint, in combination with the other
constraints, limits the maximum value of $n$ in hybrid inflation
scenarios to $n\leq 1.2$ \cite{Lyth96}.

To end, as well as summarize, this discussion,
let us rewrite Eq.~(\ref{hard}) by expressing $x_{60}^2/8\pi$
in terms of $N$ and $\Delta \phi$ by assuming that $x(\phi )$ doesn't
change too much:
\begin{equation}
(n-1) \simeq {2\over N}\ln (x_i/x_f)
        -{8\pi \over N^2}\left({\Delta\phi \over \mpl}\right)^2.
        \label{theessence}
\end{equation}
As this equation illustrates, unless $\Delta \phi /\mpl$ is
large or the steepness changes significantly, $|n-1|\la 2/N
\approx 0.04$.  This is certainly borne out by inflationary
model building:  with a few notable exceptions all models
predict $|n-1|\leq 0.1$ \cite{Lyth96}.

\section{Models with very blue spectra}

\subsection{Constraints}
The conditions for successful inflation were spelled out a
decade ago \cite{Steinh_Turner,kt}. The {\em r\`{e}gles de jeu} are:
         
$\bullet$  Slow-roll conditions must be satisfied.

$\bullet$ Sufficient number of e-folds to solve
the horizon problem ($N\ga 60$).

$\bullet$ Density perturbations of the correct amplitude
 \begin{equation} 
 \delta_H \sim V_{60}^{3/2}/V_{60}^\prime \sim 10^{-5}.
\end{equation}

$\bullet$ The distance that $\phi$ rolls in a Hubble time
 must exceed the size of quantum fluctuations, otherwise the
 semi-classical approximation breaks down
\begin{equation}
 \phidot H^{-1} \gg H/2\pi \qquad \Rightarrow \qquad V^\prime
        \gg V^{3/2}/\mpl^3,
\end{equation}
which is automatically satisfied if the density perturbations
are small.  Additionally, no aspect of inflation should hinge
upon $\phi_i$ or $\phi_f$ being smaller than $H/2\pi$, the
size of the quantum fluctuations.

$\bullet$ ``Graceful exit'' from inflation.  The
potential should have a stable minimum with zero energy
around which the field oscillates at the stage of reheating.
The reheat temperature must be sufficiently high to safely
return the Universe to a radiation-dominated phase in time
for baryogenesis and BBN.

$\bullet$ No overproduction of undesired relics such as magnetic
monopoles, gravitinos, or other nonrelativistic particles.

There are additional constraints that the potential should
obey in order to give $n\gg 1$:

(a)  $\mpl x_{60}'/4\pi$ has to be large
and positive, while $x_{60}^2/8\pi$ should be
negligible\footnote{Of course, $x_{60}^2/8\pi$ is not required to be
negligible, but it seems that it is even more difficult to get
large $(n-1)$ without this assumption.}.  Therefore $ |x_{60}|\la
O(1)$ and $\mpl x_{60}^\prime\simeq 4 \pi(n-1)$. In other words, at 60 e-folds
before the end of inflation the potential should be nearly flat and
starting to slope upwards.

(b) To obtain 60 e-folds of inflation, the potential should
be nearly flat in some region during
inflation. However, the potential must not become too flat, since then
density perturbations diverge ($\delta_H \propto 1/V^\prime$).
Therefore, the potential should have a point of approximate inflection
where $V'(\phi)$ is small but not zero.

\subsection{Example 1}

A potential with the characteristics just mentioned is
\begin{equation} V=V_0 + M^4 \left [\sinh\left (\frac{\phi-\phi_1}{f}\right )+ e^{-
\ts{\frac{\phi}{g}}}\right ],\end{equation} where $M$, $f$ $g$ and $\phi_1$ are
constants with dimension of mass.  The plot of the potential, with the
parameters calculated below, is shown in the top panel of Fig.~1.
The hyperbolic sine was invoked to satisfy
requirements (a) and (b), while the exponential
was used to produce a stable minimum.

We make the following assumptions to make the analysis simpler
(later justified by our choice of parameters below):
\vspace{.25cm}

\noindent
1)   $V_0$ dominates the potential in the slow-roll region,
\begin{equation}
 V_0 \gg M^4 \sinh \left (\frac{\phi-\phi_1}{f}\right )
\quad \mbox{for} \quad \phi_i>\phi > \phi_f. 
\end{equation}

\noindent
2) $f \gg g\;$ so that the factor $\exp(-\phi/g)$ can be completely ignored
in the slow-roll region.

\vspace{.25cm}
\noindent
3) $(\phi-\phi_1)/f$ is at least of the order of a few for
$\phi_i > \phi > \phi_f$, so that 
$\sinh [(\phi-\phi_1)/f] \gg 1$.
\vspace{0.5cm}

\noindent
4) For simplicity we take $\phi_i=\phi_{60}$.\\

In terms of the dimensionless parameter
$K \equiv \ds{\frac{M^4 \mpl}{V_0 f}}$,
\begin{eqnarray}
x &\simeq& K \cosh\left (\frac{\phi-\phi_1}{f}\right ) , \\
 x'&\simeq&  \frac{K}{f} \sinh\left (\frac{\phi-\phi_1}{f}\right ) 
-\frac{K^2}{\mpl} \cosh^2\left (\frac{\phi-\phi_1}{f}\right ) .
\end{eqnarray}
The condition that $x_{60}\la {\cal O}(1)$ becomes
\begin{equation} 
K \cosh \left (\frac{\phi_{60} -\phi_1}{f}\right )
        \la {\cal O}(1), \label{x60_negl}
\end{equation}
and the end of inflation occurs one of the slow-roll conditions
breaks down; in this case $\mpl^2 V^{\prime\prime}/V \simeq 24 \pi$, or
\begin{equation} 
\frac{\mpl K}{f} \sinh\left (\frac{\phi_f-\phi_1}{f}\right ) 
\simeq {24\pi}.
\label{infl_ends}
\end{equation}
We can now write
\begin{equation}
(n-1)\simeq
\frac{\mpl K}{4\pi f}\sinh \left( \frac{\phi_{60}-\phi_1}{f} \right) .
\label{n-1_mypot}
\end{equation} 

That inflation produces density perturbations of the correct
magnitude implies
\begin{equation}  
\sqrt{V_0}\approx 4.3 \cdot10^{-6} x_{60}\, \mpl^2. \label{V0}
\end{equation} 
The expression for the number of e-folds can be calculated
analytically.  Introducing $\alpha=(\phi-\phi_1)/f $, we have:
\begin{eqnarray}
N &=& -\frac{8 \pi}{\mpl}\int_{\phi_i}^{\phi_f} \frac{d
\phi}{x(\phi)}   \nonumber\\
  &=&  -\frac{8 \pi f}{K \mpl} \left. \tan^{-1}[ \sinh(\alpha)]
                         \: \right |_{\alpha_i}^{\alpha_f} 
  \approx \frac{8 \pi^2 f}{K \mpl}. \    \label{N_mypot}
\end{eqnarray}
In the last equality we used the fact that both $\alpha_i$ and
$|\alpha_f |$ are at least of the order of a few, so that
$\tan^{-1}[\sinh(\alpha_i)]\approx -\tan^{-1}[\sinh(\alpha_f)]\approx
\pi/2$.  This assumption will also be fully justified with our choice
of parameters below.

Finally, the potential should have a stable minimum (with $V=0$)
at some $\phi=\phi_R$. This implies that $V(\phi_R)=0$ and 
$V'(\phi_R)=0$. 

\begin{figure}[ht]
\centering 
\epsfig{file= 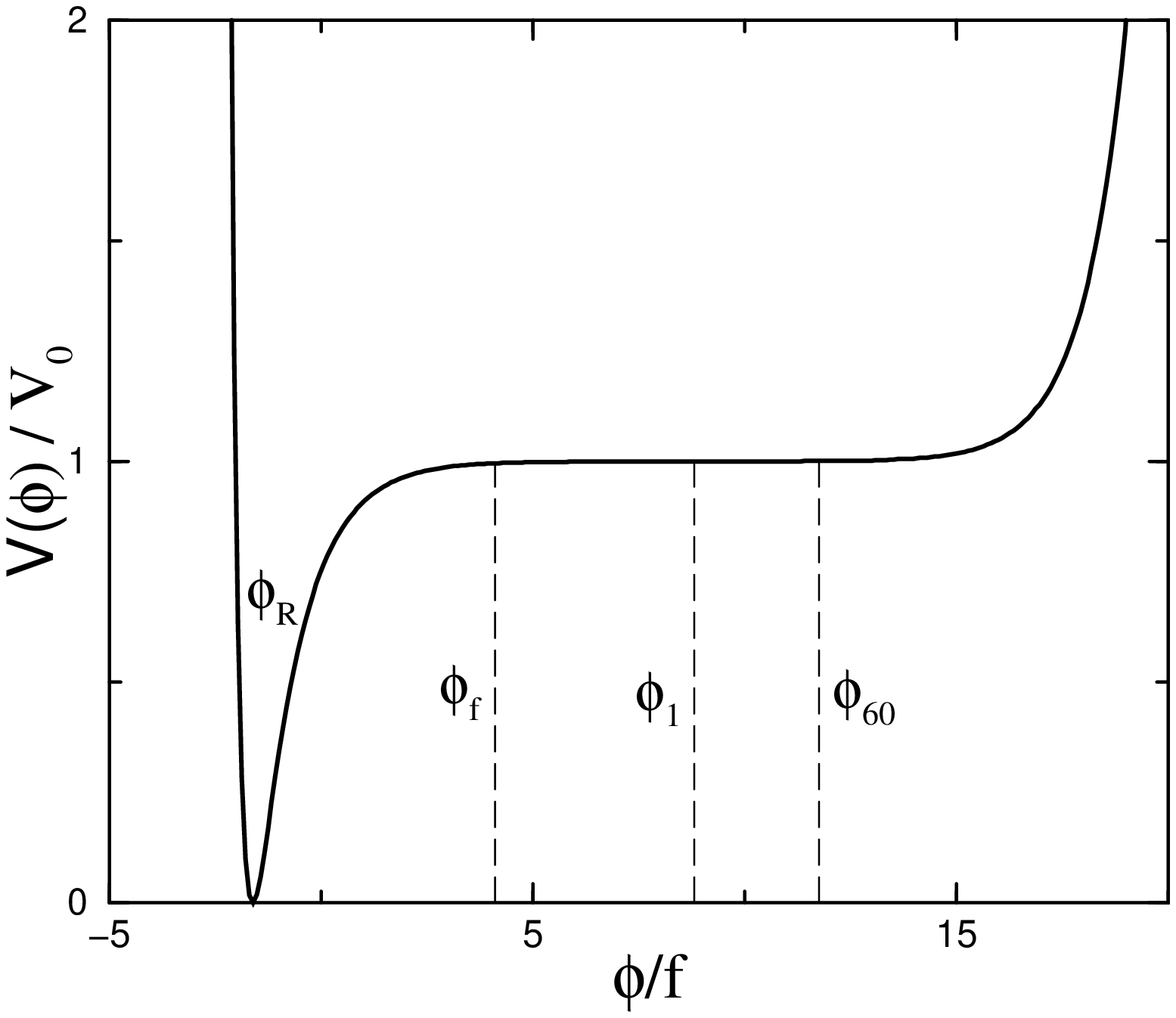, height=6.5cm, width=8cm}
\epsfig{file= 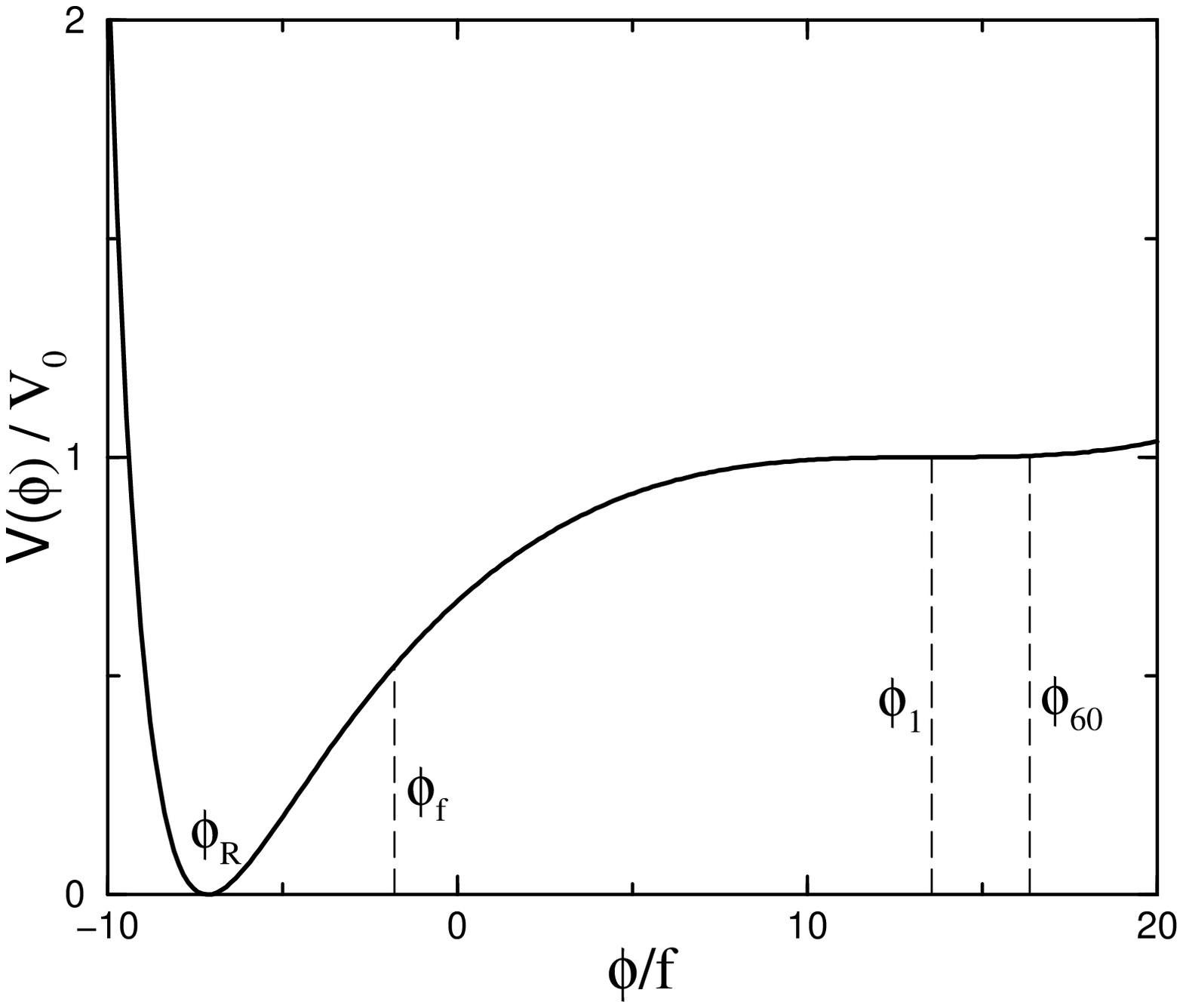, height=6.5cm, width=8cm}
\caption{Two potentials with $n\approx 2$.  In each case inflation
starts at $\phi_{60}$ and ends at $\phi_f$; the potential has a point of
approximate inflection at $\phi=\phi_1$ and its minimum at
$\phi=\phi_R$. 
Top: $V = V_0 + M^4 \left[ \sinh [(\phi-\phi_1)/f]
+ \exp(-\phi/g) \right]$. 
Bottom: $V=V_0 + M^4\left[ (\phi-\phi_1)/f + [(\phi-\phi_1)/f]^3
+ \exp(-\phi/f) \right]$.
Potential parameters are given in the text.}
\end{figure}

Before proceeding, we must specify $n$.  We choose,
somewhat arbitrarily, $n=2$. Of course, for such a large $n$ we should
include terms beyond the lowest order, complicating
the analysis. But we are
not looking for accuracy -- if $n=2$ is obtainable to first order,
then one can certainly say that $n \gg 1$ is obtainable.
(In fact, for the two potentials chosen, the second-order correction
decreases $n-1$ only slightly.)

We now have to choose parameters $V_0$, $M$, $f$, $g$, $\phi_1$,
$\phi_{60}$, $\phi_f$ and $\phi_R$ to satisfy Conditions (\ref{x60_negl} -
\ref{N_mypot}), as well as $V(\phi_R)=0$ and
$V'(\phi_R)=0$.  The choice of these parameters is by no means unique,
however.  Here is such a set:

\parbox{4.5cm}{
\begin{eqnarray*}
& V_0 = 1.7 \cdot 10^{-13}\,\mpl^4 & \\ & M^4 = 1.3\cdot 10^{-17} \,\mpl^4 &\\
& f = 7.6 \cdot  10^{-3} \,\mpl & \\ & g = f/5 &  
\end{eqnarray*}}
\parbox{2.5cm}{
\begin{eqnarray*}
&\ds{ \frac{\phi_1}{f}} = 8.80&\\ &\ds{ \frac{\phi_f}{f} } = 4.10 & 
\\ & \ds{ \frac{\phi_{60}}{f} } = 11.75. &
\end{eqnarray*}
}
\parbox{1cm}{\begin{eqnarray}\end{eqnarray}}

To verify our analytic results we integrated the equation of
motion for $\phi$ numerically and computed the spectrum
of density perturbations.  We did so
neglecting the $\ddot{\phi}$ in the
equation of motion for $\phi$ and the kinetic energy of the field
(slow-roll approximation) and taking both these quantities into
account.  The result is that $N_{\rm slow\ roll} = 57.3$ and
$N_{\rm exact}= 57.9$.  Thus, the field really rolls as predicted
by analytic methods ($N\approx 60$), and
the slow-roll approximation holds well for this potential.

\begin{figure}[ht]
\centering 
\epsfig{file= 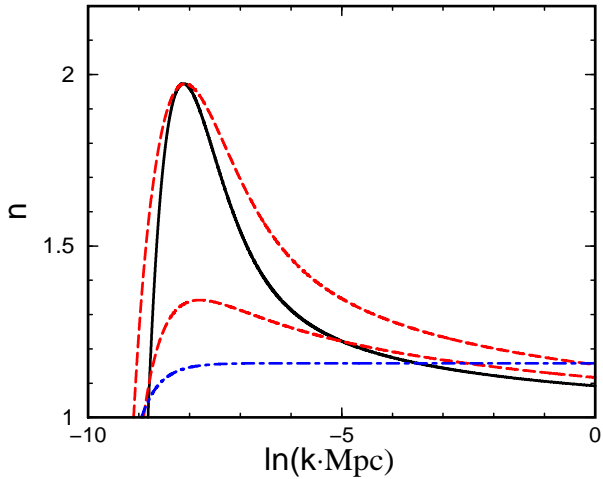, height=6.5cm, width=8cm}
\caption{The power-law index $n$ for the two inflationary potentials
constructed to give $n\sim 2$ as a function of $\ln k$.  The solid curve
corresponds to the hyperbolic sine potential and the broken curve to the
cubic potential.  While both potentials achieve $n\sim 2$, neither has a
very good power-law spectrum.  Also shown is a cubic potential model
with $n\simeq 1.4$, where the variation of $n$ is less severe.  For
comparison, the hybrid inflation model (dash-dotted curve) with $n\simeq
1.2$ is also shown; here $n$ is fairly constant over the astrophysically
interesting range.  }
\end{figure}

The numerical results for the spectrum of density perturbations
did contain a surprise, shown in Fig.~2.  While this potential
achieved large $n$, slightly smaller than 2, over a few e-folds
$n$ falls to a smaller value\footnote{Starting the roll higher on
the potential will increase the highest $n$ achieved  without violating
any of the constraints. However, $n$ will fall to equally low
values after a few e-folds as with the original $\phi_i$. }.
Indeed, even restricting the spectrum to astrophysically
interesting scales, $1\,$Mpc to $10^4\,$Mpc, the spectrum is not
a good power law, $| dn/d\ln k| \sim 0.3$, and is reminiscent of
the ``designer spectra'' with special features constructed in
Ref.~\cite{bbs}.  The reason is simple: in achieving $x^\prime
\sim 1$ an even larger value of $x^{\prime\prime}$ was attained.

\subsection{Example 2}

Is there anything special about the hyperbolic sine? Not really -- for
example, a potential of the form ``$\phi + \phi^3$'' also works.
Consider the potential
\begin{equation} 
V=V_0 + M^4 \left [\left (\frac{\phi-\phi_1}{f}\right )+
  \left (\frac{\phi-\phi_1}{f}\right )^3 + e^{-\ts{\frac{\phi}{g}}}\right].
\end{equation}
Again, we assume that $V_0$ dominates during inflation, that
$\phi_i=\phi_{60}$ and that $\exp(-\phi/g)$ can be ignored in the
inflationary region. To evaluate $N$, we further assume
that $|(\phi_{60}-\phi_1)/f| \gtrsim 1$ and $|(\phi_f-\phi_1)/f| \gtrsim
1$.  All of these assumptions are justified by the choice of parameters below.

The analysis of the inflationary constraints is similar.
We conclude that large $n$ (here $n= 2$) is
possible, with the following parameters:

\parbox{4.5cm}{
\begin{eqnarray*}
& V_0 = 1.09 \cdot 10^{-12}\,\mpl^4 & \\ & M^4 = 1.46\cdot 10^{-16} \,\mpl^4
&\\ & f = g = 1.33 \cdot 10^{-2}\,\mpl &
\end{eqnarray*}}
\parbox{2.5cm}{
\begin{eqnarray*}
&\ds{ \frac{\phi_1}{f}} = 13.54 & \\ &\ds{ \frac{\phi_f}{f} } = -1.82 &
\\ & \ds{ \frac{\phi_{60}}{f} } = 16.34. &
\end{eqnarray*}
}
\parbox{1cm}{\begin{eqnarray}\end{eqnarray}}

This potential is shown in the bottom panel of Fig.~1.
Numerical integration of the equation of motion shows that our ``60
e-folds'' is actually $N_{\rm slow roll}=55.0$ and
$N_{\rm exact}=56.0$.   Further, just as with the hyperbolic
sine potential, $n\sim 2$ is achieved, but the spectrum of
perturbations is not a good power law.  Both potentials
achieve a large change in steepness by having inflation occur
near an approximate inflection point; however, the derivative
of the change in steepness is also large, and $n$ varies significantly.
The change in $n$ can be mitigated at the expense of a smaller
value of $n$; see Fig.~2.

\section{Conclusions}\label{concl}

The deviation of inflationary density perturbations from exact
scale invariance ($n=1$) is controlled by the steepness of the
potential and the change in steepness, cf. Eq.~(\ref{n-1}).
The steepness of the potential also controls the relationship
between the amount of inflation and change in the field
driving inflation, $N\sim 8\pi (\Delta \phi /\mpl )/x$.  A very
``red spectrum'' can be achieved at the expense of a
steep potential and prolonged inflation ($t_f/t_i \gg 1$ and
$\Delta \phi \gg \mpl$); the simplest example is power-law inflation.
A very ``blue spectrum'' can be achieved at the
expense of a large change in steepness near an inflection point
in the potential and a poor power law.  In both cases
there appears to be a degree of unnaturalness.

The robustness of the inflationary prediction of that density 
perturbations  are approximately scale-invariant is  expressed  
by Eq.~(\ref{theessence}),
$$
(n-1) \simeq {2\over N}\ln (x_i/x_f)
        -{8\pi \over N^2}\left({\Delta\phi \over \mpl}\right)^2.
        \nonumber
$$
Unless the change in steepness of the potential is large,
$|\ln (x_i/x_f)| \gg 1$, or the duration of inflation is very
long, $\Delta \phi \gg \mpl$, the
deviation from scale invariance must be small, $|n-1|\la {\cal O}
(2/N) \sim 0.1$.  Even for an extreme range in $n$,
say from $n=0.5$ to $n\sim 1.5$, the variation of $\delta_H$
over astrophysically interesting scales, $\sim$1\,Mpc to
$\sim 10^4\,$Mpc, is not especially large -- a factor of $10$
or so -- but is easily measurable.

Inflation also predicts a nearly scale-invariant spectrum of
gravitational waves (tensor perturbations).  The deviation from scale
invariance is controlled solely by the first term in $(n-1)$
\cite{Turner93,Liddle_Lyth}, $n_T = -x_{60}^2/8\pi$.  Thus, only a red
spectrum is possible, with the same remarks applying as for density
(scalar) perturbations with $n\ll 1$.  In addition, the relative
amplitude of the scalar and tensor perturbations is related to the
deviation of the tensor perturbations from scale invariance, $T/S \simeq
-7n_T$ ($S$ and $T$ are respectively the scalar and tensor contributions
to the variance of the quadrupole anisotropy of the CBR).  Detection of
the gravity-wave perturbations is an important, but very challenging,
test of inflation; if, in addition, the spectral index of the tensor
perturbations can be measured, it provides a consistency test of
inflation \cite{Turner_GW}.

Finally, measurements of the anisotropy of the CBR and of the power
spectrum of inhomogeneity today which will be made over the next decade
will probe the nature of the primeval density perturbations and
determine $n$ precisely ($\sigma_n \sim 0.01$) \cite{Eisenstein_etal}.
By so doing they will provide a key test of inflation and provide
insight into the underlying dynamics.  On the basis of our work here, as
well as previous studies (see e.g., Ref.~\cite{Lyth96}), one would
expect $(n-1) \sim {\cal O}(0.1)$ or less, but not precisely zero.  The
determination that $|n-1| \ga {\cal O}(0.2)$, or for that matter $n=1$,
would point to a handful of less generic potentials.  The deviation of
$n$ from unity is a key test of inflation and provides valuable
information about the underlying potential \cite{recon}.

\end{document}